# Notícias do OPD

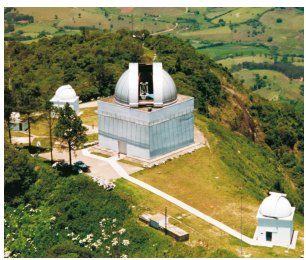

## Telescópio de 1,60 m P-E
Vinicius de Abreu Oliveira

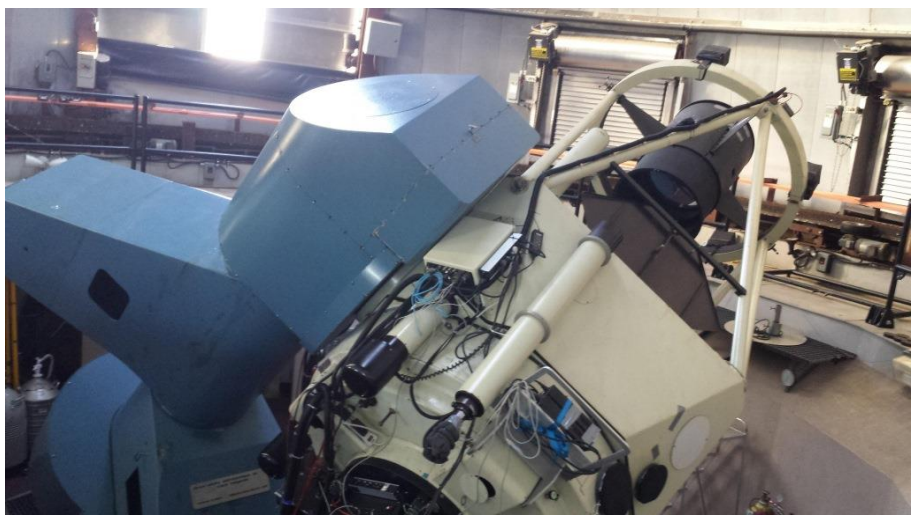

Figura 01 – Telescópio de 1,6 m de diâmetro no espelho principal, construído pela Perkin-Elmer, com projeto óptico de Ritchey-Chrétien. Imagem de abril de 2014

Com toda a certeza muitos de vocês, para não dizer todos, reconheceram o telescópio da figura 01, embora acredito que tenham achado algo de estranho. Eu mesmo tive um problema de localização espaço-temporal ao olhar de perto o nosso querido 1,6 m P-E em outro local! Sim, este é o "irmão gêmeo" que está localizado no Observatoire Mont-Mégantic (OMM), em Québec, QC, Canadá[1].

Nosso irmão é mantido pelo Centro de Pesquisa em Astrofísica de Québec (CRAQ – da sigla em francês) e administrado em conjunto pela Université Laval (na cidade de Québec) e pela Université de Montréal (na cidade de mesmo nome). Uma pequena diferença estrutural entre os irmãos é que o brasileiro tem uma razão focal de f/10, enquanto o quebequá tem uma razão focal de f/8. Este é o maior telescópio da costa Leste da América do Norte, igual nosso querido 160 cm – que é o maior da costa Leste da América do Sul.

O observatório está situado no topo do Monte Mégantic, a uma altitude de 1 111 m e a uma distância média de 250 km das cidades de Montréal e de Québec – a saber, cada uma para um lado, esta para o Norte e aquela para o Oeste. A cúpula é um pouco diferente daquela do OPD, está instalada em um prédio menor em altura e num formato mais exótico (Figura 02), alguns equipamentos são diferentes, mas a montagem e o telescópio são os mesmos. Claro que, considerando que a evolução após a construção foi completamente distinta, hoje as soluções para alguns problemas comuns são diferentes. Algo como separar fisicamente uma espécie por gerações … Darwin ficaria orgulhoso por sua teoria ser aplicada, de forma bem rude eu confesso, aos telescópios.

1- Visite o site: http://omm.craq-astro.ca/)

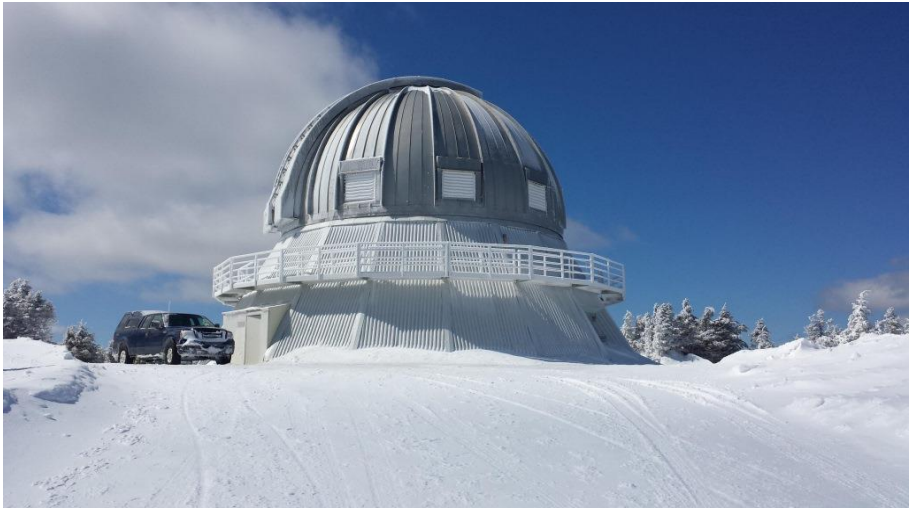

Figura 02 – Cúpula do telescópio de 1,6 m do Observatório Mont-Mégantic, prédio de dois andares com o telescópio e sala de comando ocupando o segundo andar. Imagem de abril de 2014.

A equipe responsável é bem menor do que a do OPD, são três técnicos que revezam entre si o papel de assistentes noturnos e diurnos, um diretor geral, um diretor executivo, um zelador e uma cozinheira. De forma geral, a parte administrativa é realizada pelas universidades, e a parte eletro-mecânica e de inovação é realizada em parceria com empresas quebequás de tecnologia e engenharia. Segundo me informaram, seria uma forma do governo da província incentivar o crescimento destas empresas, além de criar um pólo industrial tecnológico forte. A parceria universidade-indústria aqui é bem intensa, o que facilita o desenvolvimento de novas soluções. Por outro lado, quando o problema é algo rápido e de fácil solução, os técnicos não possuem material e autonomia para resolverem eles mesmo! Não se pode ter o melhor dos dois mundos.

Bem, de fato, quando vi o telescópio pela primeira vez, logo me vieram boas recordações, de repente tudo ficou mais familiar e o trabalho passou a ser mais fácil. Nesta missão, o "mano" está sendo utilizado com o espectroscópio SpIOMM[2] – um imageador com Transformada de Fourier, que pode ser considerado como o protótipo do SITELLE, que será utilizado pelo CFHT no Havaí (EUA). Dentre as várias vantagens deste equipamento, que não vêm ao caso neste pequeno artigo, uma das maiores é o campo de visão: algo do tipo 10' x 10'. Desta forma resulta em um cubo de dados – largura x altura x comprimento de onda – de todo o alvo, pelo menos daqueles que estou observando aqui. Para quem sempre usou fenda longa, deslocando em off-set para cubrir todo o objeto, é uma evolução muito bem vinda.

Comparando as fotos (Figura 03) se vê que as dificuldades e os anos foram gentis com ambos, mesmo com o quebequá tendo iniciado o trabalho dois anos antes do que o brasileiro, tendo a primeira luz em 1978. Sobre as soluções encontradas por cada equipe, no telescópio quebequá existem pequenas aberturas laterais próximos ao espelho principal que permitem minimizar a condensação sobre o espelho. Dificilmente observável na figura 03, mas inclusive conta com pequenas lâmpadas infravermelhas para agilizar o processo, antes de iniciar as observações. Por outro lado, o brasileiro já tem fechamento automático do telescópio!

---

2 - Para saber mais: http://www.astro.phy.ulaval.ca/staff/laurent/ARTICLES/OSA-3page.pdf

OPD

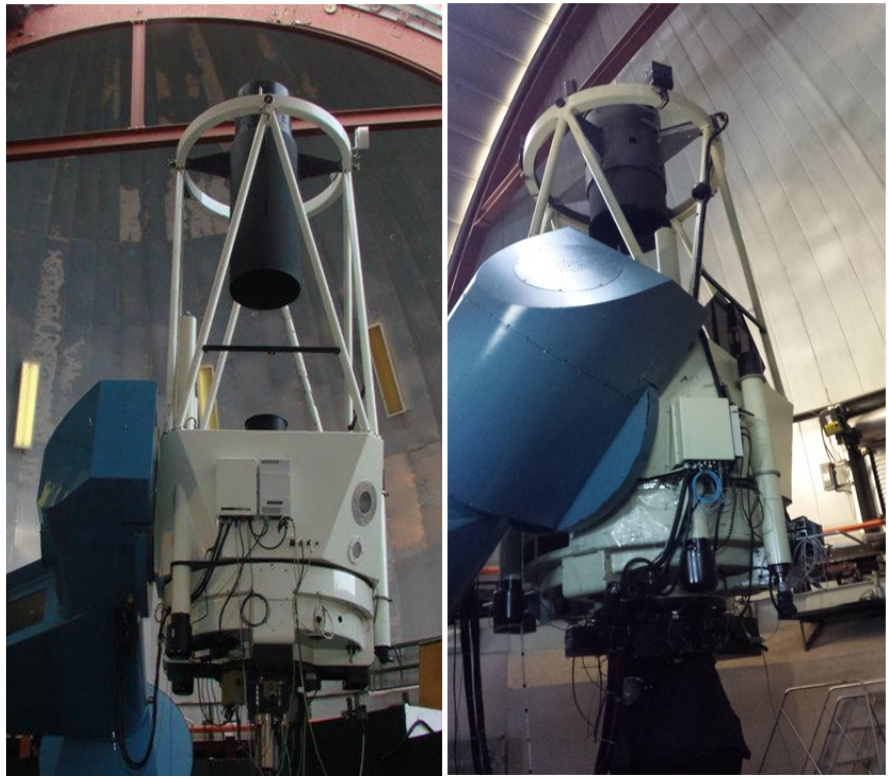

Figura 03 – A imagem a esquerda é o "mano do Sul", imagem disponível no site do OPD; a imagem da direita é o "mano do Norte" com o equipamento SpIOMM acoplado, imagem de abril de 2014

Já para uma comparação acadêmico-científica, temos o gráfico da figura 04, onde se percebe que a quantidade de tese e dissertação (produtividade acadêmica) são equivalentes em números absolutos, embora o OMM seja utilizado, basicamente, por duas universidades apenas. Com certeza isso deixa bem claro que o "mano" daqui está sendo sub-utilizado em suas funções de treinamento de novos astronômos.

Por outro lado, a quantidade de artigos científicos (produtividade científica) para o OPD é muito superior, mesmo considerando que os valores para o OPD englobam mais de um teléscópio, e não apenas o de 1,6 m de diâmetro. Aqui podemos verificar o efeito oposto, possivelmente por termos muito mais pesquisadores com ascesso direto ao OPD, em relação ao OMM. Esta característica pode ser melhor percebida ao estimarmos uma média de artigos para o intervalo de 14 anos da amostra, o que resulta em 17 artigos para o OPD e 8 para o OMM.

Por fim, vemos que os irmãos seguiram suas "vidas", formando novas gerações de astronomos em seus países sede e produzindo ciência. Esperamos que por muitos anos ainda.

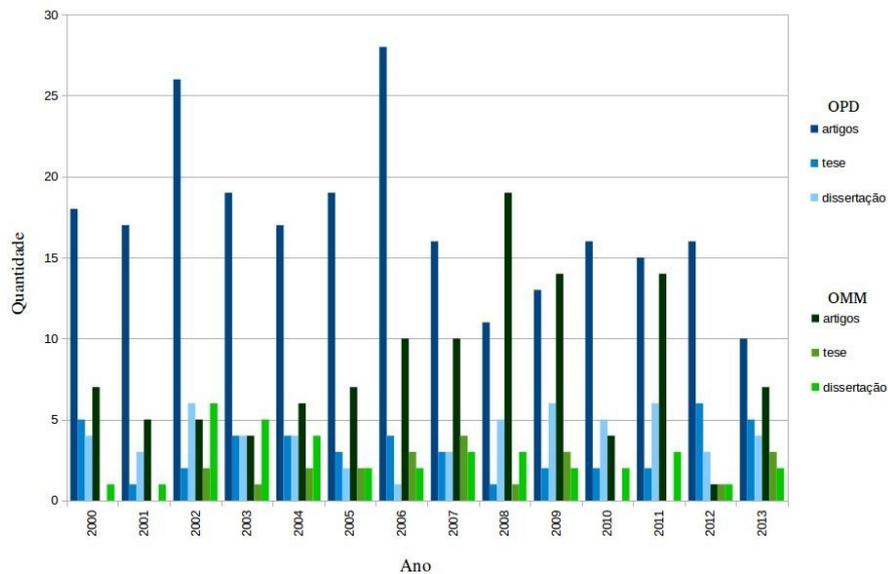

Figura 04 – Comparação entre as publicações realizadas com dados do Observatório Pico dos Dias (OPD) e do Observatoire Mont-Mégantic (OMM), para os anos entre 2000 e 2013.

Vinicius é professor adjunto na Universidade Federal do Pampa (UNIPAMPA), Campus Caçapava do Sul; atualmente está afastado para realizar pós-doutorado junto ao Grupo de Pesquisa em Astrofísica, na Université Laval, em Québec, QC, Canadá.

# Chamada para submissão de proposta observacional para o semestre 2014B

Mark Pereira dos Anjos

## Prazo final: dia 30 de abril às 24h (horário de Brasília)

A Secretaria da Comissão de Programas do Observatório do Pico dos Dias (SECOP/OPD) vem comunicar que o prazo para submissão de propostas de observações astronômicas no Observatório do Pico dos Dias já está aberto. O prazo iniciou-se no dia 1° de abril e encerra-se no dia 30 de abril às 24h (horário de Brasília).

A avaliação dos projetos é realizada com base no mérito científico, viabilidade técnica e produtividade científica dos pesquisadores envolvidos.

O encaminhamento dos pedidos de tempo é aceito somente através de Formulário on-line, disponível no endereço:

http://www.lna.br/opd/info_obs/prazos_-form.html

Dúvidas gerais sobre a submissão de projetos e formulários on-line, contacte a SECOP (opd.secop@lna.br); questões técnicas e operacionais, contacte o Coordenador do OPD, Rodrigo Campos (rodrigo@lna.br).

OPD

Mark Pereira dos Anjos é assistente em C&T e mebro da SECOP/OPD